\documentclass[reprint,superscriptaddress]{revtex4-1}
\usepackage{blindtext}
\usepackage{graphicx}
\usepackage{amsmath,amssymb}
\usepackage{multirow}
\usepackage{url}
\usepackage{color,soul}
\usepackage[colorlinks,
            linkcolor=blue,
            anchorcolor=blue,
            citecolor=blue,
            ]{hyperref}
\usepackage{setspace}

\begin{document}

\title{Experimental Realization of Anti-Unitary Wave-Chaotic Photonic Topological Insulator Graphs Showing Kramers Degeneracy and Symplectic Ensemble Statistics}

\author{Shukai Ma}
\email{skma@umd.edu}
\affiliation{Quantum Materials Center, Department of Physics, University of Maryland, College Park, Maryland 20742, USA}
\author{Steven M. Anlage}
\affiliation{Quantum Materials Center, Department of Physics, University of Maryland, College Park, Maryland 20742, USA}

\begin{abstract}

Working in analogy with topological insulators in condensed matter, photonic topological insulators (PTI) have been experimentally realized, and protected electromagnetic edge-modes have been demonstrated in such systems.
Moreover, PTI technology also emulates a synthetic spin-1/2 degree of freedom (DOF) in the reflectionless topological modes.
The spin-1/2 DOF is carried by Quantum Valley Hall (QVH) / Quantum Spin Hall (QSH) interface modes created from the bianisotropic meta waveguide (BMW) platform, and realized both in simulation and experiment.
We employ the PTI setting to build an ensemble of wave chaotic 1D metric graphs that display statistical properties consistent with Gaussian Symplectic Ensemble (GSE) statistics.
The two critical ingredients required to create a physical system in the GSE universality class, the half-integer-spin DOF and preserved time-reversal invariance, are clearly realized in the QVH/QSH interface modes.
We identify the anti-unitary T-operator for the PTI Hamiltonian underlying our experimental realization.
An ensemble of PTI-edgemode metric graphs are proposed and experimentally demonstrated. 
We then demonstrate the Kramers degeneracy of eigenmodes of the PTI-graph systems with both numerical and experimental studies.
We further conduct spectral statistical studies of the edgemode graphs and find good agreement with the GSE theoretical predictions. 
The PTI chaotic graph structures present an innovative and easily extendable platform for continued future investigation of GSE systems.

\end{abstract}

\maketitle

\section{I. Introduction}

In the 1950s, it was proposed that Random Matrix Theory (RMT) describes the energy level spacing statistics of large nuclei with various symmetries \cite{Wigner1955,Dyson1962,Mehta2004}.
Based on the BGS (Bohigas-Giannoni-Schmit) conjecture, RMT was later applied to describe the statistical properties of a broader range of the quantum/wave properties of systems that show chaos in the classical limit \cite{Haq1982,Bohigas1984,Alhassid2001}.
The Dyson three-fold way \cite{Dyson1962, RN429779} classifies the statistics of chaotic systems in three different universality classes.
Systems that show time-reversal invariance (TRI) are described by the Gaussian orthogonal ensemble (GOE), the statistics of systems showing broken time-reversal invariance are described by the Gaussian unitary ensemble (GUE), and the statistics of systems with TRI and anti-unitary symmetry show Gaussian symplectic ensemble (GSE) statistics \cite{Dyson1962, Scharf88, Mehta2004, Dutta2007, HaakeFritz2010QSoC, Beenakker2015}.
Both GOE and GUE statistics were observed in a wide variety of chaotic systems, and in particular, microwave billiard systems \cite{So1995, Stoffregen1995, Stock99, Richter2001, Hemmady2005, Dietz2015, Gradoni2014, Ma2020}. 
Complex metric graphs made with standard microwave coaxial cables have been shown to have statistical properties described by the GOE and GUE random matrix ensembles to a good approximation, although deviations are seen in some properties \cite{Hul2004, Hul2009,Biaous2016, Dietz2017, Fu17, RN429927}.

Due to the lack of a photonic analog of a spin-1/2 degree of freedom in polarization-preserving wave propagation, the realization of a GSE system was not observed for a long time in classical wave systems.
Recently, a microwave experiment utilizing a carefully designed cable network shows clear evidence of GSE statistics \cite{Joyner2014, Rehemanjiang2016}.
This work relies on a specially engineered construction involving two geometrically identical sub-graphs, along with the creation of phase shifts inside and between the sub-graphs that enables an anti-unitary symmetry for the eigenmodes.
The secular matrices of the network, written based on Kirchhoff's laws, directly resemble the form of the Hamiltonian of a spin-1/2 system, and a demonstration of the GSE energy level spacing statistics is demonstrated.
However, this method lacks generality since delicate tricks are required: the creation of two sub-graphs with exactly equal cable lengths and carefully tuned $0$ and $\pi$ phase shifts for waves traveling between the two sub-graphs \cite{Joyner2014, Rehemanjiang2016, Rehemanjiang2018, Rehemanjiang2020, Lu2020, RN429948, RN429549}. 
{As far as we know, there still awaits a physical realization that displays a clear and recognizable analog of full spin-1/2 physics in an intrinsic manner for the excitations of the system.
In particular, we seek a physical realization in which eigenfunction and other properties can be studied in a manner similar to those employed for other wave/quantum chaotic systems, such as billiards.
}

Unprecedented wave phenomena have been introduced with the creation of photonic topological insulator (PTI) technologies \cite{Ma2020pti, RN427363}.
One of these phenomena is the synthetic spin degree of freedom (DOF) emulation in the photonic context. 
In particular, a pair of topologically protected modes with opposite effective spins has emerged.
Mesoscopic condensed matter physics concepts, including the quantum Hall (QH), quantum spin Hall (QSH), and quantum valley Hall effects (QVH), have been demonstrated as classical analogs in the photonic context in a series of landmark papers \cite{Haldane2008,Hasan2010,Kraus2012, Raghu2008, Xie2018, Gao2017}. 
Researchers have emulated nontrivial topological edgemodes in many photonic structures \cite{chen2014experimental, cheng2016robust, Khanikaev2017, ma2019topologicall}.
The bi-anisotropic metawaveguide (BMW) is one type of PTI system which can host all QH, QSH, and QVH analog phases for electromagnetic waves as perturbations of one basic structure \cite{Ma2015, Xiao2016pti, Gao2017, Ma2019pti, Ma2020pti}.
The remarkable topologically-protected transmission properties of PTI have excited a wide range of real-life applications, for example, delay line devices, isolators, and circulators \cite{Ma2020pti, Xiao2016pti, Lan2022}.
However, the spin-1/2-like degree of freedom of the topological modes is not as widely exploited.
Here we aim to create a microwave realization of a symplectic Hamiltonian system by using BMW structures with both QSH and QVH domains \cite{Ma2017,Gao2017}.
Because the QSH-QVH interface waveguide has TRI and can host edgemodes with opposite synthetic spin DOFs, such a structure is an ideal building block for realizing complex structures that fall within the GSE universality class.
We have recently demonstrated a chaotic 1D graph with topologically trivial photonic crystal defect waveguides and found good agreement with the corresponding RMT theoretical predictions \cite{Ma2023pc}. 
Our next step is to consider a chaotic graph structure where the BMW QSH-QVH interface waveguides realize the graph bonds, and represent the first use of topological waveguides for wave chaos studies.

In this paper, we first present the anti-unitary T-operator for the BMW Hamiltonian, directly linked to the Kramers degeneracy.
We then introduce the design of the QSH-QVH eigenmode topological graph system.
We present the clear observation of double (Kramers) degeneracy using numerical tools from photonic band structure studies. 
Electrically large metric graphs based on QSH-QVH interface modes are then studied for their mode-spacing statistics.
We also conducted transmission measurements of the experimentally realized graph structure and find evidence of the Kramers degeneracy in the eigenmodes of the PTI graph.

\begin{figure*}
\centering
\includegraphics[width=1\textwidth]{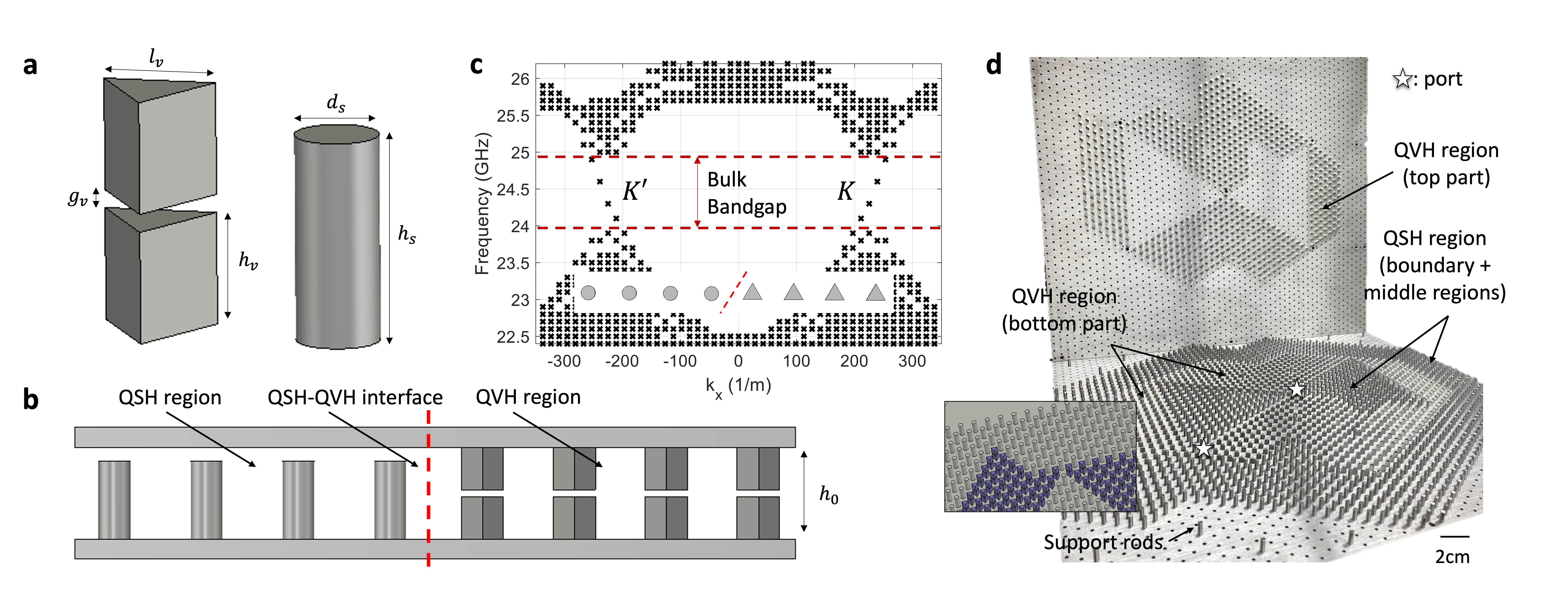}
\caption{\label{fig:PTIdesign} 
\textbf{a.} Schematic of the unit cell structure that defines the QVH and QSH regions. The quantities $l_v, g_v, h_v$ are the prism side length, the air gap distance, and the prism height of one QVH rod. The quantities $d_s$ and $h_s$ are the diameter and height of the QSH cylinder rod.
\textbf{b.} The side-view of the QSH-QVH structure with an interface (red dashed line). The top and bottom metallic plates are separated by $h_0$.
All QSH rods for a given region are attached to one of the metallic plates, and the QVH rods are mounted on both plates, leaving an air gap.
\textbf{c.} The photonic band structure of the QSH-QVH interface waveguide.
Inset shows a schematic of the supercell simulation model (same structure as in \textbf{b}). The red dashed line marks the interface between the QSH (left) and the QVH (right) region.
\textbf{d.} shows the top and bottom metallic plates and the rods/prisms attached to each one of them in the experimental realization of a GSE graph.
The top plate has the top part of the QVH rods.
On the bottom plate, we have installed the QSH rods as well as the bottom part of the QVH rods. 
Several supporting rods are installed which spread out the weight of the top plate evenly, and provide a fixed and uniform separation between the plates, $h_0$.
The inset introduces a node structure in the pyramidal graph, where the QSH and QVH rods are colored in grey and blue, respectively.
}
\end{figure*}

\section{II. Anti-Unitary T-operator}

We are interested in using the QSH-QVH interface BMW waveguides as the building block of chaotic graphs that fall in the Gaussian symplectic ensemble (GSE) universality class, appropriate for time-reversal invariant systems with a spin-1/2 degree of freedom.
An essential feature of a spin-1/2 system is an anti-unitary T-operator (satisfying $T^2=-1$).
For a time-reversal invariant system, $T^2=-1$ leads to $|\psi\rangle$ and $T|\psi\rangle$ being two different states with the same energy, creating what is known as a Kramers degeneracy.
Other than a limited number of works \cite{Wu2015,He2016,Sun2019}, the derivation of the anti-unitary condition $T^2=-1$ is not widely discussed for PTI systems, and is not discussed at all in the literature for the BMW system, to our knowledge.

As introduced before, researchers have demonstrated experimentally that the BMW system can host different topological domains in one composite structure \cite{Gao2017, Kang2018, Ma2019pti}.
The BMW structure starts with a 2D hexagonal photonic crystal (PC) sandwiched between two conducting plates.
The physical dimension of the lattice is carefully engineered so that the TE and TM modes are degenerate at the Dirac points ($K$ and $K'$ points) of the PC band structure.
At the Dirac points, a synthetic spin-1/2 DOF is emulated by the orthogonal linear combinations of the TE and TM modes.
For QSH, QH, and QVH BMW systems, the effective Hamiltonian near the Dirac points is written as \cite{Ma2015}
\begin{equation} \label{eq:H}
    \mathcal{H} = v_D\, s_0 (\tau_z \, \delta k_x \sigma_x + \tau_0 \, \delta k_y \sigma_y) +   \omega_D \Delta_{em} \tau_z \, s_z \, \sigma_z, 
\end{equation}
where $\tau$, $s$, and $\sigma$ are the Pauli matrices acting on the space of $K$ and $K'$ valley, the spin, and the Dirac degeneracy, respectively.
The quantity $\omega_D$ is the frequency at the Dirac points, and $v_D$ is the group velocity for both the degenerate TE and TM modes near the Dirac point.
Note that the Kronecker product ($s \, \sigma \equiv s \otimes \sigma$) shorthand is used.
The quantity $\delta k = (\delta k_x, \delta k_y)$ refers to the distance to the Dirac point in $k$-space.
The second term in Eqn. \ref{eq:H} represents one of several types of perturbations that can be made to the hexagonal PC lattice, in this case creating an effective spin-orbit interaction, discussed in more detail below. 
The BMW Hamiltonian has a similar form to the Kane-Mele Hamiltonian, which is used to model graphene-like topological insulators \cite{Ma2015}. 
The eigenvectors are written as $\Psi_{K (K')}^{\uparrow (\downarrow)}$, where the superscript $\uparrow (\downarrow)$ refers to the spin-up (-down) index and the subscript represents the valley index \cite{Ma2015}.
The second term introduces bianisotropy and contains $\Delta_{em}$ which is the overlap integral of unperturbed TE and TM modes (subscripts $e, m$) inside the perturbed unitcell volume.
Different topological orders are then achieved by introducing different perturbations to the basic unit cell.
For the two topological domains used in this paper, the QSH and QVH phases are realized by breaking the out-of-plane mirror symmetry, and the in-plane rotation symmetry, respectively \cite{Ma2017}. 
These symmetry-breaking perturbations will lift the mode degeneracy at the Dirac point, and create a full band gap in the PC band structure.
The detailed introduction of the BMW system Hamiltonian theory is not the goal here, and we kindly direct the readers to find more detailed discussions in Refs. \cite{Ma2015,Ma2017}.

Here we propose an effective T-operator $T_b=i\tau_x s_y \sigma_0 C$ for the BMW system, where $C$ is the complex conjugation operator ($C \, i = -i \, C$), and $\tau$ is the Pauli matrix that acts on the valley degree of freedom ($K$ and $K'$).
We want to first point out that the proposed T-operator, $T_b=i\tau_x s_y \sigma_0 C$, clearly satisfies the anti-unitary criterion $T_b^2 = -1$, leading to the Kramers degeneracy. 
The physical meaning of the operator $T_b$ lies in that it flips momentum (valley) and the spin degrees of freedom of the edgemode, expressed as
\begin{equation} \label{eq:sum}
T_b \, {\Psi}_{K (K')}^{\uparrow}={\Psi}_{K' (K)}^{\downarrow}, \quad T_b \, {\Psi}_{K (K')}^{\downarrow}=-{\Psi}_{K' (K)}^{\uparrow}.
\end{equation}
Moreover, a time reversal invariant system has $[\mathcal{H},T]=0$, or in the momentum representation $T \mathcal{H}(k) T^{-1} = \mathcal{H}(-k)$, and the proposed effective T-operator $T_b$ clearly satisfies this criterion.

\section{III. PTI-Graph design}

Recent works have reported that the BMW-PTI system can host both QSH and QVH domains inside one composite structure \cite{Gao2017,Kang2018}.
Here we present a new version of the QVH and QSH BMW-PTI structure and experimentally construct chaotic graph systems which consume $\sim 1000$ unit cells.
We designed the structure to operate around 24.5GHz to shrink down the physical size of the graph structure based on the electromagnetic scaling law.
As shown in Fig. \ref{fig:PTIdesign} (a), the QVH region unit cell consists of two identical metallic prisms with length $l_v=4.30$ mm, height $h_v = 4.32$ mm, and the middle air gap $g_v = 0.69$ mm.
Each prism is firmly attached to one of the two parallel metallic plates.
The QSH region unit cell is a metallic cylinder with diameter $d_s = 3.22$ mm and height $h_s = 8.3$ mm.
Both regions are placed between two metallic surfaces with a spacing of $h_0=9.33$ mm in the vertical direction.
The cylinder is firmly attached to one metallic plate and a gap of thickness $g_s = h_0 - h_s = 1.01$ mm is made with the second metallic plate.
The lattice constant of the triangular BMW lattice is $a_0 = h_0 = 9.33$ mm.
These physical dimensions are carefully engineered to ensure the QSH and QVH bulk bandgaps are matched in both center frequency and bandwidth.  
This bandgap occurs when the guided wavelength of waves in the interface region is comparable to $a_0$.
Fig. \ref{fig:PTIdesign} (b) illustrates how the QSH and QVH rods are mounted between the two metallic covering plates.

We examine the existence of edgemodes in the QSH-QVH interface waveguide through the so-called supercell simulation in COMSOL, as shown in Fig. \ref{fig:PTIdesign} (c).
The supercell structure consists of a slice of QSH and QVH rods (8 rods for each domain in our simulation). The Floquet boundary condition is applied to the two long sides to simulate an infinitely long waveguide structure.
It is shown that the QSH-QVH waveguide can host two counter-propagating edgemodes between $23.8 \sim 24.8$ GHz, and the two modes have opposite spin and valley DOFs.
Near the Dirac point, the synthetic spin-1/2 DOF is simulated with the plus and minus linear combinations of the TE and TM degenerate modes \cite{Ma2015}.
These propagating modes show up inside the bulk bandgap region in Fig. \ref{fig:PTIdesign} (c), with opposite propagation directions.
{
In real-space, the edgemodes and bulkmodes also show distinct eigenfunction distributions. 
The edgemodes reside strictly at the interface of different topological domains (marked by the red dashed line in} Fig. \ref{fig:PTIdesign} (b) and inset of (c){), and the mode amplitude decays exponentially into the surrounding domain regions.
The eigenfunction distribution of the bulkmodes reside in the interior of one or more topological domains. 
The clear distinction in the mode eigenfunction is another indicator for distinguishing between edgemodes and bulkmodes.}

The field of quantum chaos originated from nuclear physics studies, where the spectral statistics of the Hamiltonian of large nuclei are found to be well described by the statistics of the eigenvalues of random matrices \cite{Dyson1962, Mehta2004, RN2947, RN24757, RN427913}.
It was later conjectured that all complicated quantum/wave systems that have chaotic classical/ray counterparts possess universal statistical properties described by RMT \cite{Bohigas1984}.
Such chaotic phenomena have been manifested in wave systems with different dimensionalities \cite{Gradoni2014, Dietz2015, Ma2020, Ma2020b, Ma2022}.
The quantum graph system is an inter-connected network of transmission lines that meet at nodes/junctions \cite{Kottos1997, RN429429, Kaplan2001}. 
Waves propagate on the network according to the telegrapher equations which resemble the Schrodinger equation, and the conservation of probability current imposes a Neumann boundary condition at each node.
Previous work has shown that photonic crystal defect waveguides with trivial topology can be used as quantum graphs, and display many of the same statistical properties of more conventional cable graphs \cite{Ma2023pc}.
The proposed PTI symplectic quantum graph structure can be considered a square pyramid realized in the 2D plane (see the inset in Fig. \ref{fig:simu}(c)).
Each bond of the graph is built with a QSH-QVH interface waveguide.
The interface waveguides meet at nodes with valance 2 or 4.
The graph will be electrically large (i.e. each bond will be many wavelengths long) and create multiple paths between any two nodes, so that the scattering will be complex.

The open-plate view of the QSH-QVH PTI-graph experimental realization is shown in Fig. \ref{fig:PTIdesign} (d).
Because of the relatively high operating frequency ($\sim 24.5$ GHz), we can create a graph structure with 60 lattice periods on a side but with moderate overall size, on the scale of 50 cm $\times$ 50 cm.
To prevent the bending and flexure of the covering plates, we have also installed several support rods (with diameter $d_s$ and height $h_0$) at several locations in the horizontal plane in the bulk regions, away from the graph bonds.

The closed graph has numerous eigenfrequencies and corresponding eigenfunctions.  
When the graph is connected to one or more outside leads at discrete ports, one can define a frequency-dependent scattering matrix $S$.  
The ports are installed at the graph nodes where multiple bonds meet, and each port consists of a coaxial cable mounted perpendicular to one plate whose center conductor extends $\sim 2$ mm into the graph below the top metal plate.
The scattering matrix is measured as a function of frequency using a Keysight N5242A vector network analyzer.

An ensemble of PTI graphs can be realized by changing each graph bond's length, similar to studies of topologically-trivial PC \cite{Ma2023pc} and waveguide-based \cite{RN429258} graphs.
The losses for edge modes propagating in the structure are dominated by conductor loss in the plates, prisms, and cylinders, and their points of contact with the plates.  
A small amount of radiation loss also exists out of the 4 open edges of the finite-size structure.
The conductive loss of the structure can be approximated by a uniform attenuation $\eta$, and its value can be reduced by cooling down the graph from room temperature using a dilution refrigerator.

\section{IV. Numerical simulation studies}

\begin{figure*}
\centering
\includegraphics[width=0.95\textwidth]{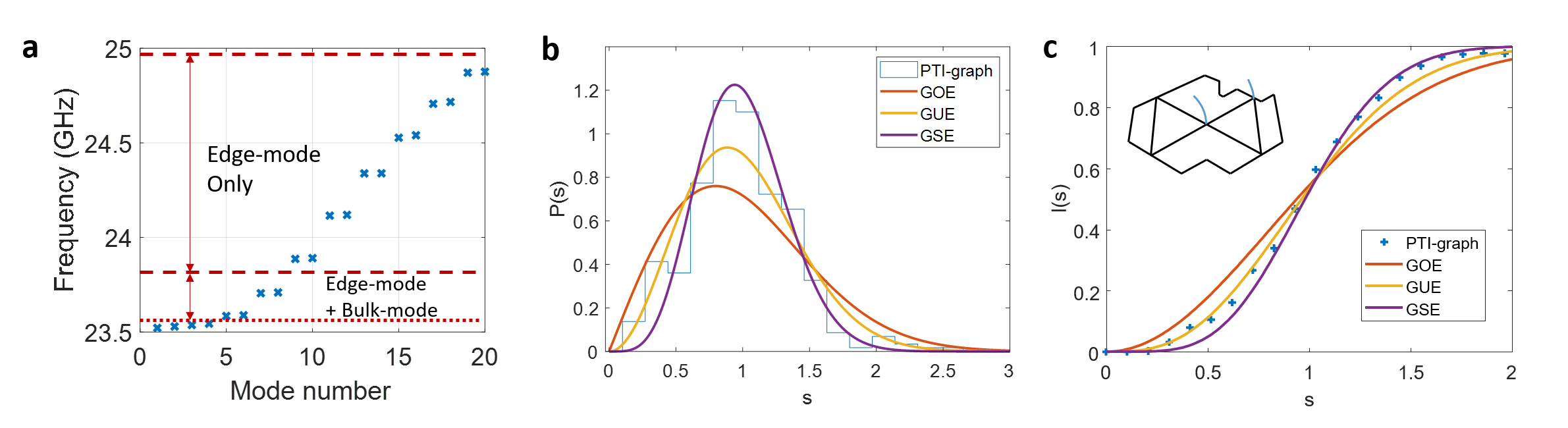}
\caption{\label{fig:simu} 
\textbf{a.} The simulated eigenmode results of a closed PTI-graph realization.
It is clearly shown that the two-fold degeneracy is present at the bulk bandgap frequencies.
At the edge of the bandgap ($\sim 23.5$ GHz), we observe a hybridization of the edgemode and bulk modes.
\textbf{b.} The histogram approximation probability distribution of the PTI graph nearest-neighbor mode-spacing $s$. The prediction of the GOE, GUE, and GSE ensembles are plotted as solid lines for reference. 
\textbf{c.} The integrated probability function $I(s)$ along with the theory predictions of the three groups.
The inset shows the schematic of the square pyramidal graph.
The black and blue lines represent the graph bonds and cables connected to the ports, respectively.
}
\end{figure*}

We first present the eigenmode study of closed PTI graphs using the numerical simulation software (CST). 
The simulated graphs have zero uniform loss, although loss is present in the simulation due to leakage through the boundaries of the finite-size simulation domain, where radiating boundary conditions are assumed.
Inside the bulk bandgap frequency range ($24 \sim 24.8$ GHz as shown in Fig. \ref{fig:PTIdesign} (b)), we observe only doubly degenerated edgemodes, as shown in Fig. \ref{fig:simu} (a).
At the lower edge of the bulk bandgap ($\sim 23.5$ GHz), a co-existence of the degenerate edgemodes and the higher spectral-density bulkmodes is observed, where the spectrum becomes nearly continuous and the doubly degenerate modes mix with the bulk modes \cite{key}.
Thus the double (Kramers) degeneracy of the QVH-QSH edge modes is clearly demonstrated with numerical tools.

We next study the nearest neighbor level spacing statistics of the proposed pyramidal graphs. 
{The nearest neighbor level spacing distribution is an excellent diagnostic of the universality class that governs the fluctuations present in the properties of a wave chaotic system} \cite{Dyson1962, Mehta2004, Stock99, HaakeFritz2010QSoC}.
The eigenvalue simulations are conducted using the software HFSS running on the UMD DeepThought2 HPC cluster.
The simulated graph has the area of $\sim 35\times 60$ photonic crystal lattice constants, where $\sim 12$ levels (2 degenerate modes per level) are found in the topological edgemode bandwidth.
For the nearest neighbor level spacing study, the quantity $s_i = e_{i+1} - e_i$, where $e_i$ is the graph eigenmode level and the subscript $i$ refers to the mode index.
The quantity $s_i$ is then normalized as $s_i/\langle s_i \rangle$ where $\langle s_i \rangle$ is the averaged level spacing.
We simulate 30 different configurations of the PTI graph and obtain 342 level-spacing values.
In Fig. \ref{fig:simu} (c) inset we show one configuration of the pyramidal graph.
New graph configurations are created by varying the lengths of graph bonds, which is implemented by adding/removing zig-zag paths on graph bonds.
Due to the topological protection, the waveguide modes can travel through these zig-zag paths without back-scattering.
By varying the length of the bonds, we are able to change the spectrum of resonant modes of the entire graph structure.

{In the quantum chaos community, it is of interest to study the statistics of the system level-spacing $s$ through its distribution function, $P(s)$, and its integrated value $I(s) = \int_0^s P(s') ds'$.}
{By comparing the functions $P(s)$ and $I(s)$ with the theoretical predictions of GOE, GUE, and GSE systems, one is able to classify the symmetry group of the system of interest} \cite{Dyson1962, Mehta2004, Stock99}.
As shown in Fig. \ref{fig:simu} (b) and (c), the level spacing statistics of the pyramidal graph agree reasonably well with the GSE statistics at larger $s$ values ($s>0.5$), and show a distinctly different shape as compared to the GOE and GUE systems, especially in terms of the peak and tail of $P(s)$.
For small spacing values ($s < 0.5$), we find that the graph system deviates from the GSE class prediction.
We attribute this to the small statistical ensemble ($\sim 12$ levels per realization), and the proximity of many bulk modes to the bandgap.
We have examined the statistics of nearest neighbor spacings in finite-size GOE matrices in which alternating eigenvalues are ignored to replicate GSE statistics, as demonstrated before \cite{Mehta2004, Alt1998}.  
We find that taking ten $30 \times 30$ GOE matrices, yielding 14 nearest-neighbor pairs, show statistical properties similar to our data (see details in Supp. Mat.). 

\begin{table}
\begin{ruledtabular}
\begin{tabular}{l l l l l l}
  & PTI-Graph & GSE & GUE & GOE & PC-Graph\\
  \hline
$\left< r \right>$ & 1.05 & 1.18 & 1.37 & 1.75 & 1.89\\
$\left< \Tilde{r} \right>$ & 0.89 & 0.68 & 0.60 & 0.54 & 0.51\\
\end{tabular}
\end{ruledtabular}
\caption{\label{tab:catalog} 
The ensemble-averaged consecutive mode spacing ratios of the photonic topological insulator graph (PTI-Graph) system, the theoretical predictions for the GOE, GUE, and GSE systems \cite{Atas2013}, and the mode spacing ratios of the topologically-trivial photonic crystal graph (PC-Graph) \cite{Ma2023pc}. }
\end{table}

To augment the mode spacing distribution test, we also investigated the method of consecutive mode spacing ratios, where one computes two parameters $r_i = \frac{s_i}{s_{i-1}}$ and $\Tilde{r}_i = min \left( r_i, \frac{1}{r_i} \right)$ \cite{Atas2013}.
The mode spacing ratio ensemble-averaged data are reported in Table. \ref{tab:catalog}.
For the PTI graph system, the averaged values of $\left< r \right> = 1.05$ and $\left< \Tilde{r} \right> = 0.89$ are closer to the GSE theoretical predictions than either the GUE or GOE predictions.
Note that we also include the mode spacing ratio values of the GOE graph system consisting of topologically-trivial photonic crystal waveguides in the last column \cite{Ma2023pc}.
It is clear that a GOE graph system mode spacing ratios are at the other extreme of values, and closer to the GOE theoretical prediction values.
From this mode spacing study of numerical realizations of the PTI graph system we conclude that it is more consistent with GSE statistics than any other Gaussian ensemble.

\section{V. Experimental studies}

With the QSH-QVH waveguide carefully designed by means of numerical simulations, we fabricated a physical realization of the chaotic PTI-graph structure, as shown in Fig. \ref{fig:PTIdesign} (d). 
The PTI rods and the two covering flat plates are made of aluminum 1070 alloy (aluminum $> 99.7\%$ and iron $< 0.25\%$). 
The structure is designed to be compatible with low-temperature measurements in a dilution refrigerator (DR) so that the entire graph can undergo uniform changes in temperature to systematically vary the uniform attenuation $\eta$ of the edgemodes.
We have installed numerous support rods to prevent the drooping/bending of the covering plates to ensure uniform air gaps throughout the structure. 
The structure is measured in a vertical geometry to minimize the effects of gravity on the graph properties.
Note that the center frequency and the width of the bandgap are sensitive to the set of parameters $h_0$, $g_v$, and $g_s$, and the triangular prism QVH rods on the two cover plates need to be carefully aligned when assembling the graph structure.
It should be noted that small variations in $h_0$, $g_v$, and $g_s$, and the small variations in the orientation of the QVH rods, have no effect on the topological properties of the PTI-guided waves as long as they do not close the bandgap \cite{Kang2020}.

We conduct 2-port S-parameter measurements using a Keysight N5242A PNA-X microwave network analyzer.
Electric dipole antennas are installed at two of the graph nodes (shown as the stars in Fig. \ref{fig:PTIdesign} (d)).
The electric field inside the graph structure is coupled to the dipoles which allows one to probe the eigenmodes of the closed graphs.
The lengths of the dipoles are chosen to be short so that the coupling is weak, aiming to minimize the perturbation to the modes.
For room-temperature S-parameter measurements, the microwave network analyzer calibration planes are at the end of the dipole antenna input terminals.
For low-temperature tests, we conduct un-calibrated S-parameter measurements because the cables inside the DR experience varying temperatures during the measurement and the calibrations cannot be performed at each temperature.

Our main objective in this experiment is to identify the Kramers degeneracy expected in the QSH-QVH interface graph modes.
However, the degenerate modes are difficult to identify by simply studying the resonances in the S-parameter spectrum.
Several methods to identify degenerate modes in microwave resonant systems have been developed in the past.
For example, the perturbations caused by the ports used to measure a billiard can be used to identify two degenerate modes \cite{Kuhl2000}.
By exciting a microwave cavity at two locations with a phase-shifted signal, one can also identify degenerate modes \cite{Dembowski2003}.
In the graph structure with anti-unitary symmetry \cite{Joyner2014, Rehemanjiang2016}, the system can be perturbed away from the interference condition that underlies anti-unitarity, and observe the splitting of the Kramers degenerate modes.  
However, the topologically-protected nature of the QVH-QSH edge modes prevents such studies here.

To better examine the question of degenerate modes in the data, we compute the complex Wigner-Smith time-delay \cite{RN429395, RN428220, RN428678, Chen2021} $\tau_W = -\frac{i}{M} \frac{d}{df} ln[det(S(f))]$, which can be used to identify the location of the scattering matrix poles and zeros in the complex frequency plane \cite{Chen2021, Chen2022}.
The quantity $M$ is the number of measurement channels, and $S(f)$ refers to the full $M\times M$ complex scattering matrix, measured as a function of frequency.
We adopt the Heidelberg picture of scattering theory \cite{RN429570, RN1212, RN428721}.  
In this picture, we regard the closed graph to be described by a Hamiltonian $H$, having real eigenvalues that describe the resonant frequencies of the isolated graph.  
Coupling of the graph to the outside world through $M$ ports introduces a coupling matrix $W$ whose matrix elements $W_{i,j}$ denote the coupling of mode $i$ to port $j$.  
This creates an effective (non-Hermitian) Hamiltonian $H_{eff} = H -i \Gamma_W$, where $\Gamma_W = \pi W W^{+}$. 
The complex eigenvalues of $H_{eff}$ are the poles of the scattering matrix, written as $f_n -i \Gamma_n$, with the convention that $\Gamma_n > 0$ for passive lossy systems.  
Because the losses are uniform in our system, we shall assume that the zeros of the S-matrix are simply the complex conjugates of the poles.
The complex time delay $\tau_W$ around an isolated group of $N$ modes suffering uniform attenuation $\eta > 0$, as a function of frequency, can be fit by the following relationships \cite{Chen2021}:
\begin{widetext}
\begin{equation} \label{eq:retauw}
Re(\tau_W) = C + \frac{1}{M} \sum_{n=1}^N \left[ \frac{\Gamma_n - \eta}{(f-f_n)^2 + (\Gamma_n - \eta)^2} + \frac{\Gamma_n + \eta}{(f-f_n)^2 + (\Gamma_n + \eta)^2} \right] \end{equation}
\begin{equation} \label{eq:imtauw}
Im(\tau_W) = \frac{1}{M} \sum_{n=1}^N \left[ (f-f_n) \left( \frac{1}{(f-f_n)^2 + (\Gamma_n + \eta)^2} - \frac{1}{(f-f_n)^2 + (\Gamma_n - \eta)^2} \right) \right] \end{equation}
\end{widetext}
where the quantities $\eta, f_n -i \Gamma_n$ and $C$ are the uniform loss rate, the $n^{th}$ system complex resonant frequency (S-matrix pole), and a constant term to account for contributions to $Re(\tau_W)$ from far-away resonances, respectively.
For a single isolated resonance mode in the PTI-graph $\tau_W$ spectrum, we expect $N=2$ because there are 2 degenerate modes for each energy level.
Note that we have simplified the fitting process by assuming the zeros and poles of the S-matrix, written as $z_n$ and $f_n - i\Gamma_n$ in Ref. \cite{Chen2021}, where $n$ is the mode index, are complex conjugates of each other (i.e. $z_n = f_n + i \Gamma_n$).
This assumption is suitable for the PTI-graph systems because it has only uniform loss and there is no lumped loss element present in the graphs \cite{Chen2021, Chen2022}.  
The value of $\eta$ is dictated by resistive losses in the Aluminum plates and rods, and can be considered independent of frequency over the range of fits to Eqs. \ref{eq:retauw} and \ref{eq:imtauw}, in the bandgap of the PTI graph structures.

\begin{figure*}
\centering
\includegraphics[width=0.95\textwidth]{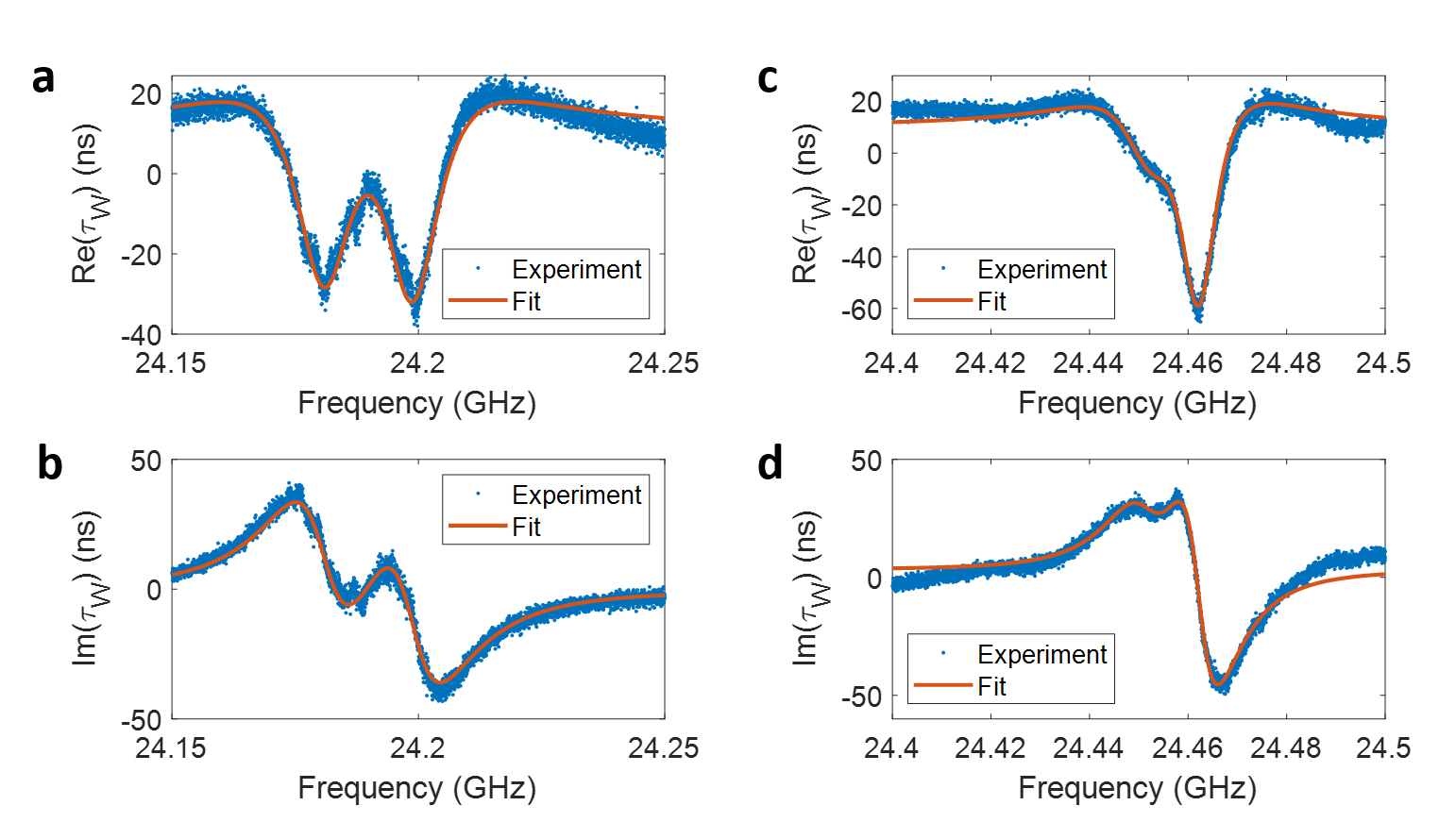}
\caption{\label{fig:roomT}
The real and imaginary parts of the measured complex Wigner time delay as a function of frequency for nearly degenerate pairs of modes around 24.20 and 24.46 GHz at room temperature, shown in (\textbf{a}, \textbf{b}) and (\textbf{c}, \textbf{d}), respectively. The fitting results are shown in solid red curves, and the fit parameters are given in the text.
The deviations between data and the fits at the extreme edges of (a), (c), and (d) are due to the presence of neighboring levels, which are not included in the fit.
}
\end{figure*}

Based on Eqns. \ref{eq:retauw} and \ref{eq:imtauw}, we are able to fit the complex Wigner-Smith time delay $\tau_W(f)$ obtained from the scattering matrix data for the PTI eigenmodes.
Two examples of the measured (blue dots) and the fitted (solid red line) complex $\tau_W$ curves are shown in Fig. \ref{fig:roomT} for an isolated degenerate mode of the PTI graph.
Note that these $\tau_W(f)$ curves have an asymmetric shape as a function of frequency.
From the forms of Eqns. \ref{eq:retauw} and \ref{eq:imtauw}, it is clear that the asymmetric $\tau_W(f)$ curves must be the result of two extremely nearby modes.
To further prove that the nearby modes are degenerate, we conduct high-quality numerical fitting of the measured data.
The best-fit parameters are $(f_1, f_2) = (24.18, 24.20)$ GHz and $(\Gamma_1, \Gamma_2, \eta) = (0.0070, 0.0073, 0.0146)$ GHz for the data in Fig. \ref{fig:roomT} (a) and (b), and $(f_1, f_2) = (24.45, 24.46)$ GHz and $(\Gamma_1, \Gamma_2, \eta) = (0.0019, 0.005, 0.0098)$ GHz for Fig. \ref{fig:roomT} (c) and (d).
One may argue that the two nearby modes shown in Figs. \ref{fig:roomT} (a) and (b) (or (c) and (d)) might just be two closely located but unrelated modes.
However, the spacing between these two modes is on the order of $\sim 10$ MHz, which is much smaller than the $\sim 70$ MHz mean mode-spacing for a graph of this size.
The mean-mode-spacing is computed by $\frac{c}{2L_{tot}}=0.073GHz$, where $c$ is the speed of light and $L_{tot}=2.05m$ is the total length of all graph bonds.
Note that the imaginary parts of the poles, as well as the uniform attenuation, are all much less than the mean spacing between edgemodes, allowing for a definitive determination of all the scattering singularities.
Thus, in the case of some of the modes, we believe that we have identified the expected 2-fold Kramers degeneracy through the measured complex time-delay spectrum.

\begin{figure*}
\centering
\includegraphics[width=0.95\textwidth]{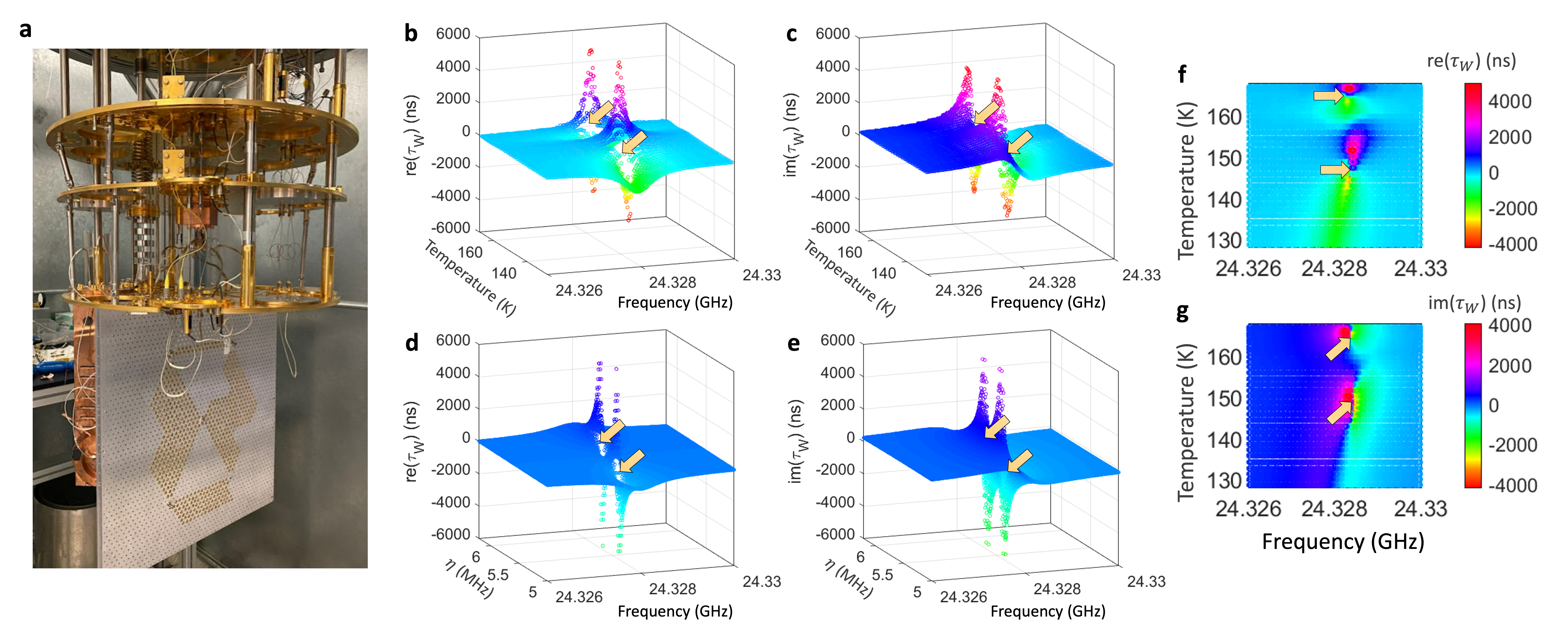}
\caption{\label{fig:lowT} 
\textbf{a} shows the front view of the closed PTI-graph structure mounted to the DR mixing chamber plate.
\textbf{b} and \textbf{c} show the real and imaginary part of the measured PTI-graph complex Wigner time delay $\tau_W$ versus temperature and frequency, respectively.
\textbf{d} and \textbf{e} show the real and imaginary part of a numeric example of complex Wigner time delay $\tau_W$ versus varying uniform attenuation $\eta$ and frequency, respectively.
The yellow arrows mark the location of $\tau_W$ divergence in \textbf{b} through \textbf{g}.
\textbf{f} and \textbf{g} show top views of the data in \textbf{b} and \textbf{c}, plotted on the same color scale.
}
\end{figure*}

For most resonant modes of the graph, the $f_n$ values of the degenerate mode pair appear to be almost identical so that the corresponding $\tau_W(f)$ curves do not have an asymmetric shape.
However, in these situations, one can utilize the difference in the imaginary parts of the S-matrix poles ($\Gamma_n$) to reveal the 2-fold degeneracy as follows.  
Two nearly-degenerate S-matrix poles at $f_1 - i\Gamma_1$ and $f_2 - i\Gamma_2$ can have $f_1 \approx f_2$ but still have $\Gamma_1 \neq \Gamma_2$.  
The difference in imaginary parts is due to the different overlap of the modes with the measurement ports, and this difference is a generic feature of chaotic graphs (note that this assumption is the basis of the Random Coupling model for the statistical properties of complex scattering systems \cite{Hemmady2005, Zheng2006, Zheng2006a, Gradoni2014}).
To find nearly degenerate modes, we shall exploit the properties of complex time delay (Eqns. \ref{eq:retauw} and \ref{eq:imtauw}) by smoothly sweeping the uniform loss $\eta$ value so that it sequentially passes through the values of both $\Gamma_1$ and $\Gamma_2$.
For example, one may strategically set the initial value (room temperature) of $\eta$ to be greater than $max[\Gamma_1, \Gamma_2]$, and then systematically decrease $\eta$ until $\eta < min[\Gamma_1, \Gamma_2]$.
During this process, the value of $\eta$ will first equal the $\Gamma_n$ value of one of the two degenerate modes ($max[\Gamma_1, \Gamma_2]$).
When this happens, the $\tau_W(f)$ spectrum will diverge as one of the denominators of Eqs. \ref{eq:retauw} and \ref{eq:imtauw} go to zero when $f=f_n$.
When $\tau_W(f)$ diverges, the $Re(\tau_W)$ will change sign and the $Im(\tau_W)$ will show high absolute values just above and below resonance.
Moreover, because one generically has $\Gamma_1 \neq \Gamma_2$, one expects to find that the $\tau_W(f)$ curve `diverges' twice during the process of systematically varying $\eta$.
We call this process a \textit{Lazarus time-delay behavior}, where varying $\eta$ has made $\tau_W$ `die'verge twice to demonstrate the presence of an eigenmode degeneracy.
This double-time-delay divergence will not take place if only one mode exists in the frequency range of interest.
Hence, cooling the graph in a dilution refrigerator (DR) from room temperature to low-temperatures will allow us to gradually decrease the degree of uniform loss $\eta$ by controlling the PTI structure global conductivity loss, while keeping the values of $\Gamma_1$ and $\Gamma_2$ approximately constant since they are dictated by the structure of the port.
In this way, we can reveal the existence of nearly degenerate modes from the evolution of complex time delay as $\eta$ is systematically varied by means of the temperature variation of the PTI graph.

To identify Kramers degenerate pairs of modes, we adopted the Lazarus method experimentally by cooling the graph from room temperature to $\sim 30$ mK.
The conductivity loss of the system, represented by $\eta$ in Eqns. \ref{eq:retauw} and \ref{eq:imtauw}, will gradually and monotonically decrease during the cooldown process \cite{RN26167}.
The coupling at $M=2$ ports is made weak enough such that $max[\Gamma_1,\Gamma_2] < \eta$ at room temperature (this is the case for both of the modes shown in Fig. \ref{fig:roomT}). 
The graph structure is suspended vertically below the mixing chamber plate (mxc plate) of a BlueFors XLD-400 DR, as shown in Fig. \ref{fig:lowT} (a).
The graph structure takes up nearly the entire diameter of the mxc plate and is a large thermal mass attached to the DR.
We used braided copper wire straps (not shown) to aid the thermal connection between the graph and the mxc plate, and to make the temperature of the entire structure uniform.
The residual resistivity ratio of Aluminum typically exceeds $10^3$ \cite{RN26167} so a reduction of $\eta$ by at least a factor of 30 is expected between room temperature and the base temperature of the DR.
Note that the S-parameter measurement is not calibrated in this case, but this does not affect our ability to identify degenerate modes from complex time delays.

Here we present the analysis of measured $\tau_W(f)$ curves during the cooldown.
Figures. \ref{fig:lowT} (b) and (c) show the measured $\tau_W(f)$ spectrum for a single pair of modes when the temperature changes from 180 K to 80 K.
As discussed above, we expect a double divergence of the $\tau_W(f)$ as the temperature-dependent value of $\eta$ is swept through $\Gamma_1$ and $\Gamma_2$.
To better illustrate the evolution of $\tau_W$ under uniform loss variation, we also present a numerical example of similar character in Fig. \ref{fig:lowT} (d) and (e).
The numerically-generated $\tau_W(f)$ data in Figs. \ref{fig:lowT} (d) and (e) is obtained using Eqs. \ref{eq:retauw} and \ref{eq:imtauw}, with parameters $(f_1, f_2) = (24.382, 24.382)$ GHz, $(\Gamma_1, \Gamma_2) = (0.0054, 0.0057)$ GHz, and $\eta$ varying from 0.0062 to 0.0050 GHz.
As shown in both the measurement and the numerical example, when $\tau_W$ diverges, we observe that $Re(\tau_W)$ changes sign and $Im(\tau_W)$ shows high magnitudes near resonance.
A clear resemblance between the measurement and the numerical example is found.
In Figs. \ref{fig:lowT} (f) and (g) we show the top view of the real and imaginary parts of the measured $\tau_W$ (same data as in (b) and (c)).
It is clear that only one dispersing feature of enhanced $\tau_W(f)$ exists throughout the measurement, as opposed to two distinct dispersing features, as one would expect for accidentally degenerate modes.
This observation means that the frequencies of the two degenerate modes remain essentially identical as the PTI graph contracts and becomes less lossy through cooling.
No other singular features in $\tau_W(f)$ are observed from this set of degenerate modes outside of this temperature range.
Figs. \ref{fig:lowT} (b) to (e) show exactly the case where the 2-fold degeneracy can not be identified through a difference in the real part of S-matrix poles/zeros ($f_n$), but can be identified from the difference in imaginary parts ($\Gamma_n$) during uniform loss ($\eta$) variation.
Effectively, we identify the presence of degenerate modes by sweeping 'imaginary frequency' through the S-matrix zeros.
We note in passing that the divergence of $Re[\tau_W]$ is an indication that an S-matrix zero has been brought to the real frequency axis \cite{Chen2021} and that coherent perfect absorption can be achieved by exciting the system with the corresponding eigenvector \cite{RN426383, Chen2020}.

\section{Discussion}

{
Through the use of complex time delay and temperature variation of the uniform attenuation of the graph, we have identified the Kramers degeneracy in the proposed PTI graph structure.  
Our work presents a clear experimental method for identifying Kramers degeneracy within PTI systems. 
Additionally, our innovative use of complex time delay and variable uniform attenuation, leading to what we term ‘Lazarus time-delay behavior’, can potentially be applied to other fields of study. 
We also note that in addition to the demonstrated method of uniform attenuation tuning, the manipulation of other parameters, such as coupling strength, can lead to similar ‘Lazarus’ time delay divergent events. Future exploration of Kramers degeneracy in other contexts is thus enabled by our experiment.

The introduction of Kramers degeneracy to microwave and optical resonant systems plays a pivotal role in extending the advantages of PTI systems.  
The presence of Kramers degeneracy within PTI systems is intricately related to the system’s band structure, endowing the modes with topological protection as long as the bandgap remains intact. 
This remarkable property enables PTI modes to travel seamlessly through sharply bent waveguides and propagate without reflection, making PTI systems invaluable for a variety of applications} \cite{Ma2020pti}.

{
The QVH/QSH PTI graph incorporates the spin-1/2 degree of freedom directly into the wave excitation that propagates throughout the system.  
This is in contrast to previous GSE experimental studies which relied on connecting two GUE graphs in a particular way to create a composite system that had anti-unitary symmetry.  
As a result of the simpler and distributed nature of our GSE realization, our approach also enables studies of many other wave chaotic properties of GSE systems, such as eigenfunction statistics, for the first time.  
This will enable a complete investigation of the statistical properties of systems making up Dyson’s three-fold way}  \cite{Dyson1962, RN429779}.

\section{Conclusion}

To conclude, we have proposed a PTI-graph structure as a new platform for realizing wave chaotic systems that display symplectic Hamiltonian behavior in a natural and general manner.
We present an effective time-reversal operator $T_b$ for the BMW system and show that it is anti-unitary.
A new composite QSH-QVH BMW interface structure is designed, fabricated, and employed as the building block for creating chaotic graphs in the Gaussian symplectic ensemble universality class.
We applied numerical tools to study the proposed graphs and find clear evidence of Kramers degeneracy with eigenvalue simulations of the closed graph structure.
We studied nearest-neighbor level spacing statistics by simulating an ensemble of such graphs.
A relatively good agreement between the GSE prediction and the graph system data is found.
We then realized an equivalent QSH-QVH BMW interface graph structure in Aluminum.
In the experimental test, we identified the Kramers degeneracy through the asymmetric shape of the measured complex Wigner-Smith time delay $\tau_W(f)$ spectrum.
With the help of variable-temperature measurements, we systematically varied the system uniform loss $\eta$ and found clear double divergence in the measured $\tau_W(f)$, demonstrating the presence of two degenerate modes.
Thus, theoretical, numerical, and experimental studies have accumulated clear evidence for Kramers degeneracy in this unique setting.
Compared to the reflection-less waveguide transmission in PTI, its delicate emulation of a spin-1/2-like DOF of light is highly under-utilized in scientific explorations \cite{Ma2020pti}.
This work serves as an example of how the synthetic spin DOF can be employed to address a long-standing puzzle in another field of physics, namely quantum chaos.

\section{Acknowledgements}
The authors thank Gennady Shvets, Yang Yu, and Ran Gladstein Gladstone for fruitful discussions.
We thank Lei Chen for discussions on the complex Wigner time delay analysis.
We thank the University of Maryland for providing access to the HPC cluster DeepThought2.
This work was supported by ONR under Grant No. N000141912481, Grant No. N000142312507, DARPA Grant HR00112120021, ONR/DURIP FY21 under grant N000142112924, ONR/DURIP FY22 under grant N000142212263, and the Maryland Quantum Materials Center.


\begin{thebibliography}{10}
\bibitem{key}
The CST-simulated edge and bulk modes are clearly distinguished by the eigenmode electric field distributions, which are either confined to the QVH/QSH interface, or leak substantially into the bulk.
\end{thebibliography}
\end{document}